\documentclass[twocolumn,conference]{IEEEtran}
\usepackage{fancyhdr}
\IEEEoverridecommandlockouts
\usepackage[stable]{footmisc}
\usepackage{graphicx}
\usepackage{amsmath}
\usepackage{mathtools}
\usepackage{algorithm}
\usepackage{algpseudocode}
\usepackage[group-separator={,}]{siunitx}
\usepackage{xspace}
\usepackage{booktabs}
\usepackage{tabu}
\usepackage{tabularx}
\usepackage{tabularray}
\usepackage{graphbox}
\usepackage{diagbox}
\usepackage{threeparttable}
\usepackage{stfloats}
\usepackage{wrapfig}
\usepackage{stmaryrd}
\usepackage{caption}
\usepackage{subcaption}
\usepackage{bbding}
\usepackage{multirow}
\usepackage{tikz}
\usepackage{pifont}
\usepackage{pgfkeys}
\usepackage{enumitem}
\usepackage{hyperref}
\usepackage{xurl}
\usepackage{cleveref}
\usepackage[framemethod=tikz]{mdframed}
\usepackage{extarrows}

\usetikzlibrary{decorations.pathreplacing}
\usetikzlibrary{shapes.multipart}
\usetikzlibrary{graphs}
\usetikzlibrary{graphs.standard}
\UseTblrLibrary{booktabs}
\usepackage{enumitem}
\setlist[itemize]{leftmargin=1em}

\begin{document}
\definecolor{gray}{rgb}{0.7,0.7,0.7}
\definecolor{darkgray}{rgb}{0.5,0.5,0.5}

\newcommand{\cred}{\textcolor{red}}
\newcommand{\cblue}{\textcolor{blue}}

\def\myfw{Mat2\-Bound\-ary}
\newcommand{\plainfwname}{\myfw\xspace}
\newcommand{\frameworkname}{\textsc{\myfw}\xspace}

\def\mtosfw{Mat2\-Stencil}
\newcommand{\plainmtos}{\mtosfw\xspace}
\title{\huge \plainfwname: Treating User-Defined Boundary Condition as SpMV for Distributed PDE Solvers on Block-Structured Grids}

\author{
    \IEEEauthorblockN{Yanzheng Cai, Mingzhe Zhang, Shengqi Chen, Haoyuan Song, Wenguang Chen$^\ast$}
    \IEEEauthorblockA{Department of Computer Science and Technology \& BRNist, Tsinghua University, Beijing, China.}
    \IEEEauthorblockA{\{cyz22,zmz21,csq20,song-hy22\}@mails.tsinghua.edu.cn, cwg@tsinghua.edu.cn}
}

\maketitle

\begingroup\renewcommand\thefootnote{$\ast$}
\footnotetext{Corresponding author.}
\endgroup

\begin{abstract}
Boundary-condition (BC) handling is a major source of complexity in PDE solvers on structured and block-structured grids, especially for high-order methods and distributed-memory execution. We present \frameworkname, a DSL and compiler for boundary computations that models a broad class of BCs as affine sparse linear operators. 
This abstraction unifies halo copying, circular and symmetric mappings, zero padding, block-edge synchronization, and user-defined interpolation, while exposing a modular basic sub-matrix interface for declarative composition. 
To make this representation efficient, \frameworkname combines multi-stage programming and polyhedral analysis to generate matrix-free kernels for structured cases, support user-defined sparse matrices for irregular cases, eliminate redundant boundary work, and synthesize reusable communication schedules for distributed execution. 
Evaluated on two shallow-water equation solvers on cubed-sphere grids and HPCG, \frameworkname achieves up to 7.6$\times$ BC-kernel speedup, reduces BC code by over 70\%, and scales to 1,344 CPU cores with 72\%-88\% efficiency.
\end{abstract}


\section{Introduction}
\label{sec:intro}
\begin{figure}[ht]
    \centering
    \begin{subfigure}[t]{0.23\textwidth}
        \centering
        \includegraphics[width=1.0\textwidth]{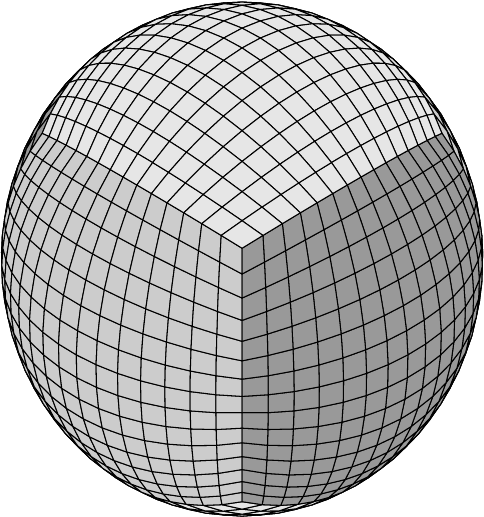}
        \subcaption{A 3D view of the cubed-sphere tiling with resolution of 15 $\times$ 15.}
        \label{fig:intro-example:cube}
    \end{subfigure}
    \hfill
    \begin{subfigure}[t]{0.23\textwidth}
        \centering
        \includegraphics[width=1.0\textwidth]{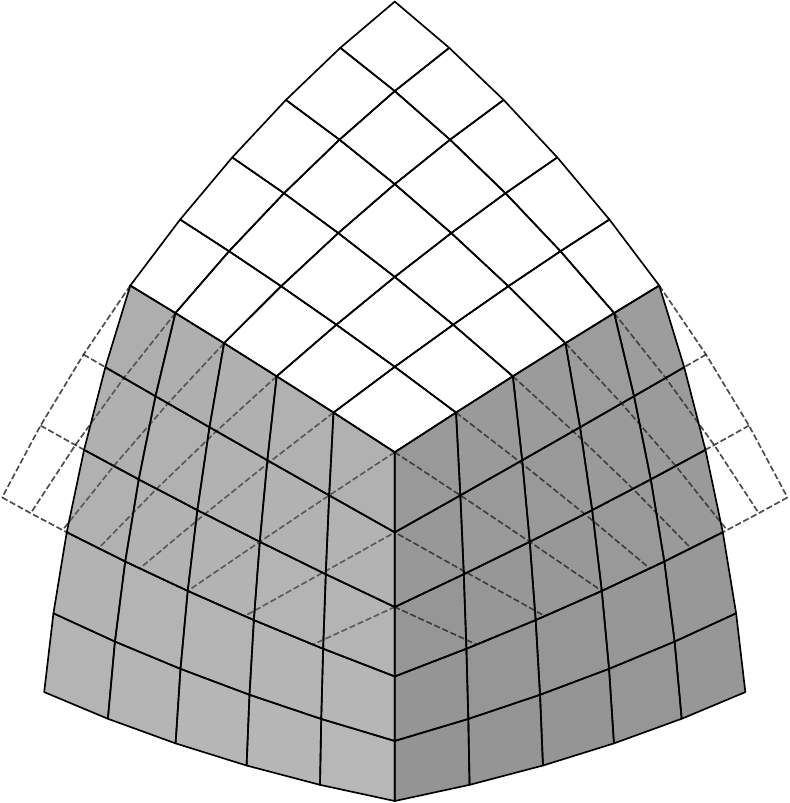}
        \subcaption{ A closeup view showing the overlap of grid lines from the upper block ghost cells on the neighboring blocks.}
        \label{fig:intro-example:corner}
    \end{subfigure}
    \caption{Geometry of the cubed-sphere grid.}
    \label{fig:intro-example}
\end{figure}

Numerical solutions of partial differential equations (PDEs) are of great importance in a wide range of scientific fields, 
including seismic imaging~\cite{seismic}, weather prediction~\cite{wrf}, etc. 
However, manually developing PDE solvers using general-purpose programming languages (e.g., FORTRAN, C/C++) imposes a substantial programming burden, requiring the implementation of thousands to tens of thousands of lines of code.
Many of these PDE solvers employ structured grids, primarily Cartesian grids, for spatial discretizations. Due to the statically determined neighboring relationships of the grid points, numerous programming frameworks and domain-specific languages (DSLs)~\cite{cao2023mat2stencil, paredes2023gt4py, devito, lengauer2020exastencils, falgout2002hypre} offer simplified programming interfaces and dedicated optimization passes for applications on structured grids 
Block-structured grids, which are the concatenations of several structured blocks, are powerful when processing complex and real geometries by offering uniform or quasi-uniform grid cells. They are widely used in real world applications, such as the dynamical core of weather and climate models~\cite{ji2016dynamical}. 
\Cref{fig:intro-example} demonstrates the cubed-sphere grid for discretizations of the sphere geometry, and we can see obvious discontinuity over block boundaries. 

While maintaining the computing efficiency and accuracy as structured grids in the center region of each block, block-structured grids involve additional boundary condition (BC) computations. 
Simple and naive implementations of the boundary algorithm can introduce degradation on the order of accuracy and grid imprinting~\cite{fv3-boundary}. Therefore, researchers have proposed several high-order interpolation schemes for the boundary calculation of cubed-sphere grids~\cite{chen2008mcv, ullrich2010high, HOPE, GungHo}. 
The complexity in the programming of these boundary algorithms is reflected in the length of the code. As shown in 
\Cref{tab:LOC}, different boundary processing logics are of 1.12-9.87x in code length compared to the corresponding inner calculation kernels, and the total BC lines of code (LOCs) account for over 40\% of the whole standalone application. 
The ratio will increase sharply when we apply program representations with much more simplicity to inner calculations on structured grids ~\cite{devito, lengauer2020exastencils,cao2023mat2stencil}, which can decrease their code length to less than 20\% of origin~\cite{cao2023mat2stencil}. 
Furthermore, process-to-process communications should also be considered when we distribute these standalone applications to large clusters, making the boundary conditions the most lengthy and tedious part of the application programs. 
Previous frameworks~\cite{paredes2023gt4py, ben2022productive} and applications ~\cite{HOPE} have tried to give solutions to different instances of boundary algorithms, but they fail to balance versatility, programmability and scalability. 
\begin{table*}[t]
    \centering
    \begin{tblr}{colspec={Q[c,m] Q[c,m] Q[c,m] Q[c,m] Q[c,m] Q[c,m]}}
        \toprule
         {Method}&  Operator&  {BC type} & {BC interp./quadrature \\dimensions} &  {BC LOCs} & {Calc LOCs} \\
         \midrule
         \SetCell[r=3]{c} {HOPE ~\cite{HOPE}}
         &  {Reconstruction} & iterative & 2d/2d & 385  & 39 \\
         &  Border Flux & direct & -/- & 178 & 159 \\
         & Other & - & -/- & - & 542 \\
         \midrule
         \SetCell[r=3]{c} {MCV ~\cite{chen2008mcv}} &  {Flux Derivative} & direct & 1d/-  & 203 & 49\\ 
         & {Result Averaging} & direct & -/ - & 168 & 23 \\
         & Other& -& -/- & - & 317\\ 
         \bottomrule
    \end{tblr}
    \caption{Statistics of coding length for implementing two standalone PDE solvers on cubed-sphere grids using FORTRAN}
    \label{tab:LOC}
\end{table*}

To address these challenges, we introduce a novel DSL called \frameworkname{}, designed for processing user-defined BCs on distributed structured and block-structured grids. 
Our contributions are as follows: 
\begin{itemize}
    \item We formalize a wide spectrum of boundary algorithms—from simple padding to complex user-defined interpolations—under a generalized SpMV abstraction, enabling modular construction via programmable basic sub-matrices.
    \item By integrating multi-stage programming with polyhedral analysis, our compiler automatically resolves boundary dependencies at compile-time, eliminates redundant iteration spaces, and generates highly optimized, matrix-free kernels for structured sub-regions.
    \item For large-scale execution, we introduce a domain-specific communication backend for low-overhead remote data fetching.  
    \textit{IR-reusing} techniques are also introduced to achieve moderate compilation time and output code length. 
    \item Integrated with a standalone dsl which describes the computation on structured grids~\cite{cao2023mat2stencil, m2s_ae}, our dsl is capable of expressing a much wider range of boundary conditions which appeared in real applications, including different high order finite-volume method (FVM) solvers for solving shallow-water equation (SWE) on cubed-sphere grids and High Performance Conjugate Gradients (HPCG) on structural grids. 
    For BC computations, we achieve a performance gain of 0.93-7.64$\times$ compared to original FORTRAN/C++ implementations with only 30\% LOCs for different SWE solvers. 
    By changing several lines of code, we can distribute these application on multiple processes, reaching an average 4.75$\times$ speed-up for standalone execution compared with original implementations,  and a strong scaling efficiency of 72\%-88\% on a cluster with 1,344 CPU cores. 
\end{itemize}

\section{Background and Motivation}
\label{sec:bg}

\subsection{Survey on Existing Boundary Algorithms}
\begin{figure}[ht]
    \centering
    \begin{subfigure}[t]{0.32\linewidth}
        \centering
        \includegraphics[width=\textwidth]{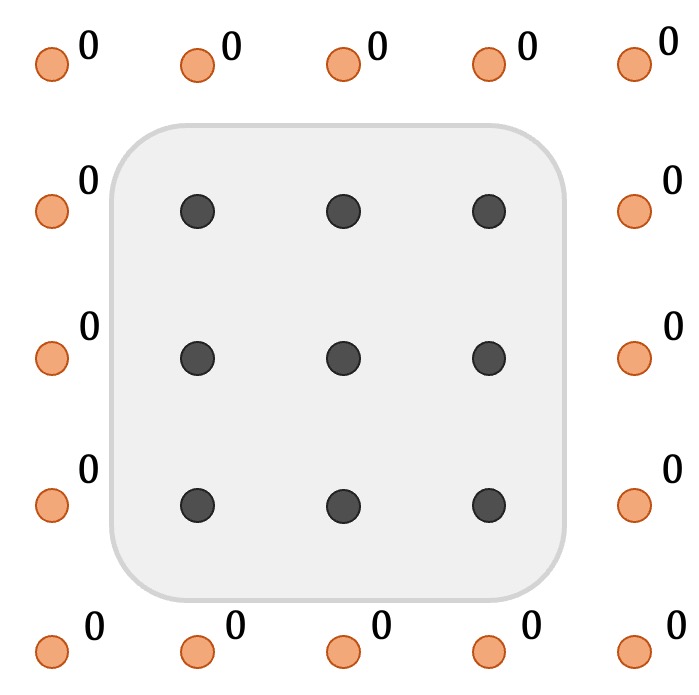}
        \subcaption{Padding Zero}
        \label{fig:bc-example:zero}
    \end{subfigure}
    \hfill
    \begin{subfigure}[t]{0.64\linewidth}
        \centering
        \includegraphics[width=\linewidth]{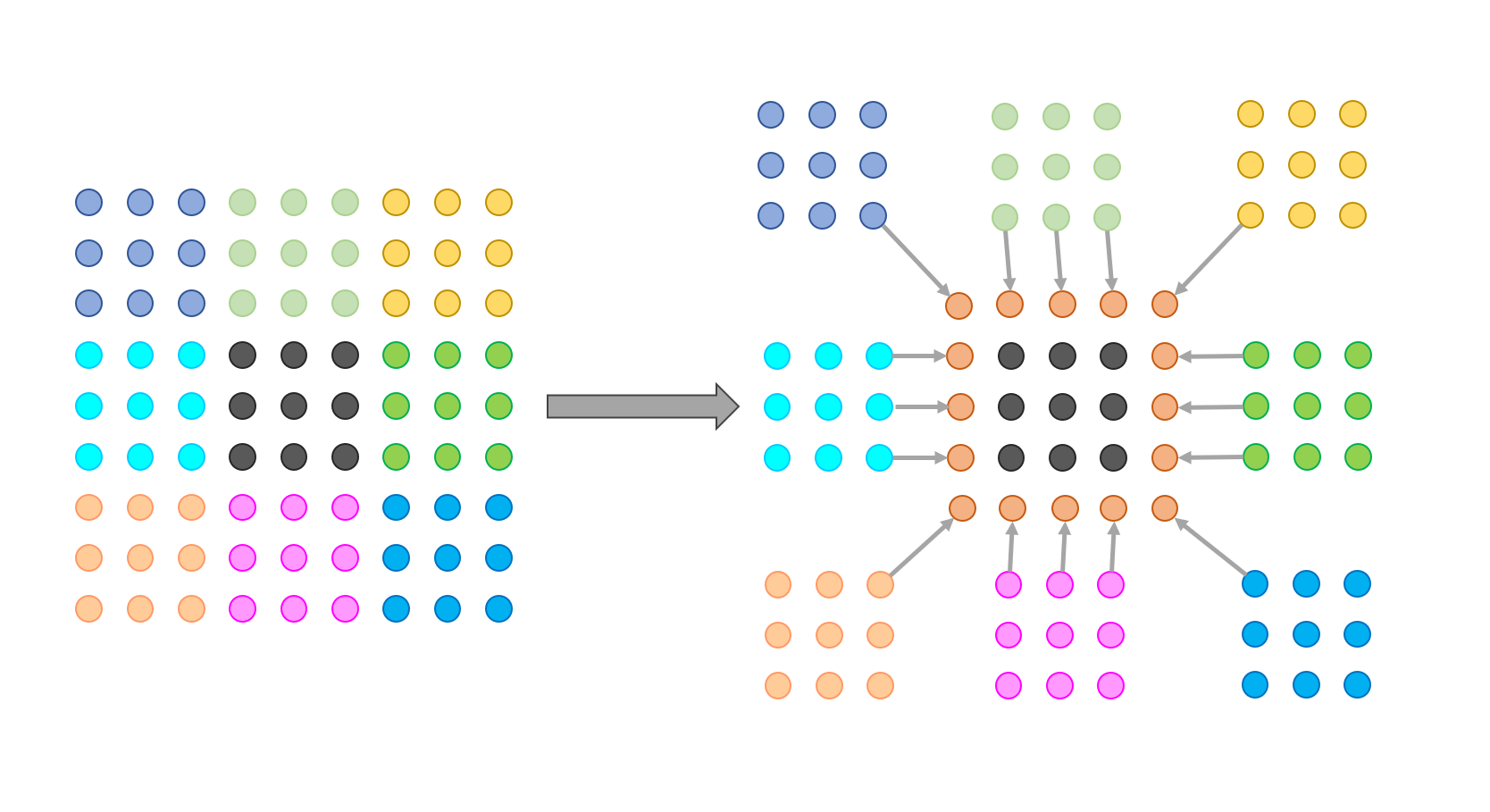}
        \subcaption{Halo Copying}
        \label{fig:bc-example:halo}
    \end{subfigure}
    \begin{subfigure}[t]{0.40\linewidth}
    \begin{subfigure}[t]{0.9\linewidth}
        \centering
        \includegraphics[width=\textwidth]{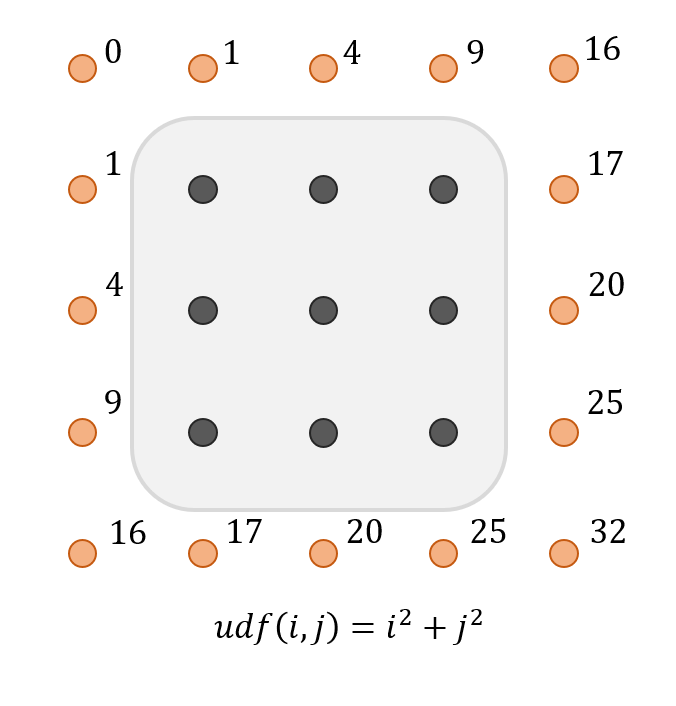}
        \subcaption{Pure Function}
        \label{fig:bc-example:pf}
    \end{subfigure}
    
    \begin{subfigure}[t]{0.9\linewidth}
        \includegraphics[width=\textwidth]{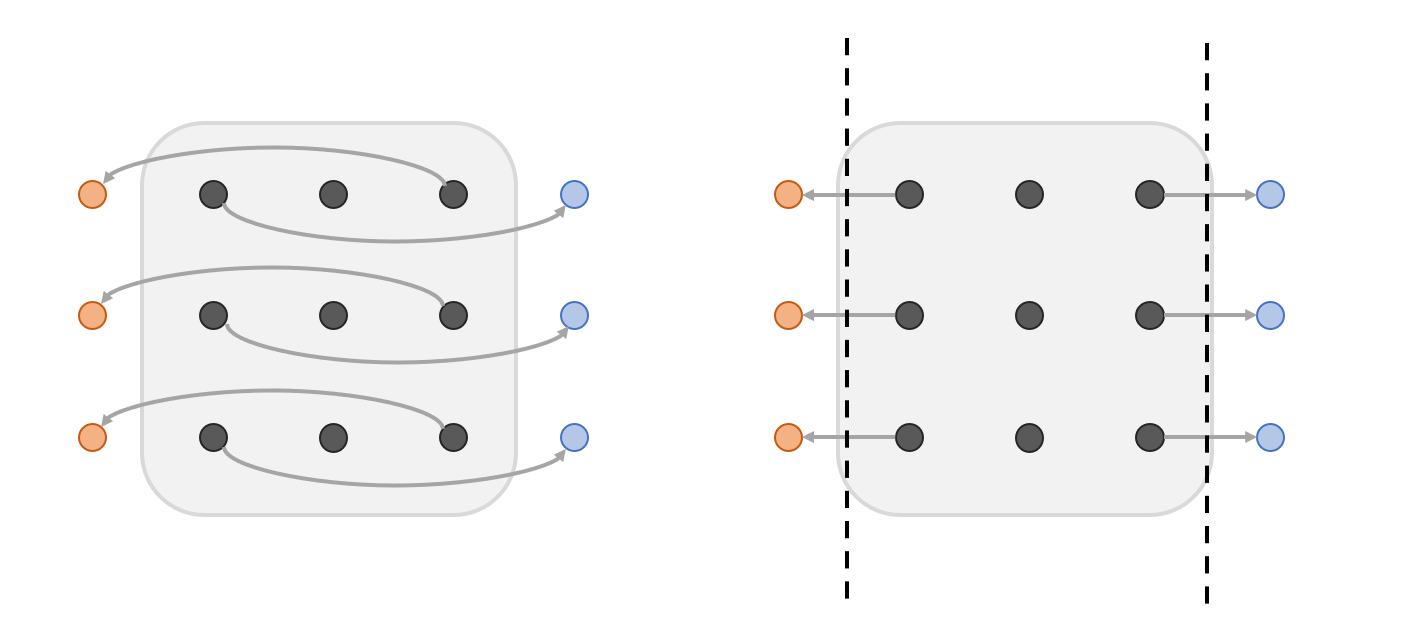}
        \subcaption{Circular/Symmetric}
        \label{fig:bc-example:cs}
    \end{subfigure}
    
    \begin{subfigure}[t]{0.9\linewidth}
        \centering
        \includegraphics[width=\textwidth]{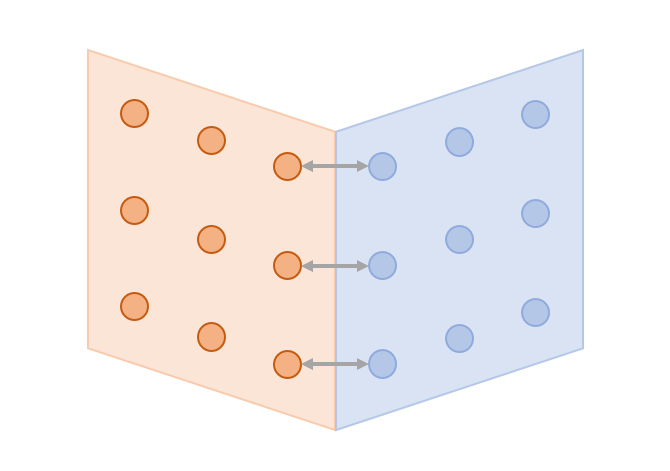}
        \subcaption{Edge Averaging}
        \label{fig:bc-example:avg}
    \end{subfigure}
    
    \end{subfigure}
    \hfill
    \begin{subfigure}[t]{0.56\linewidth}
        \centering
        \begin{subfigure}[t]{0.8\linewidth}
            \centering
            \includegraphics[width=0.9\textwidth]{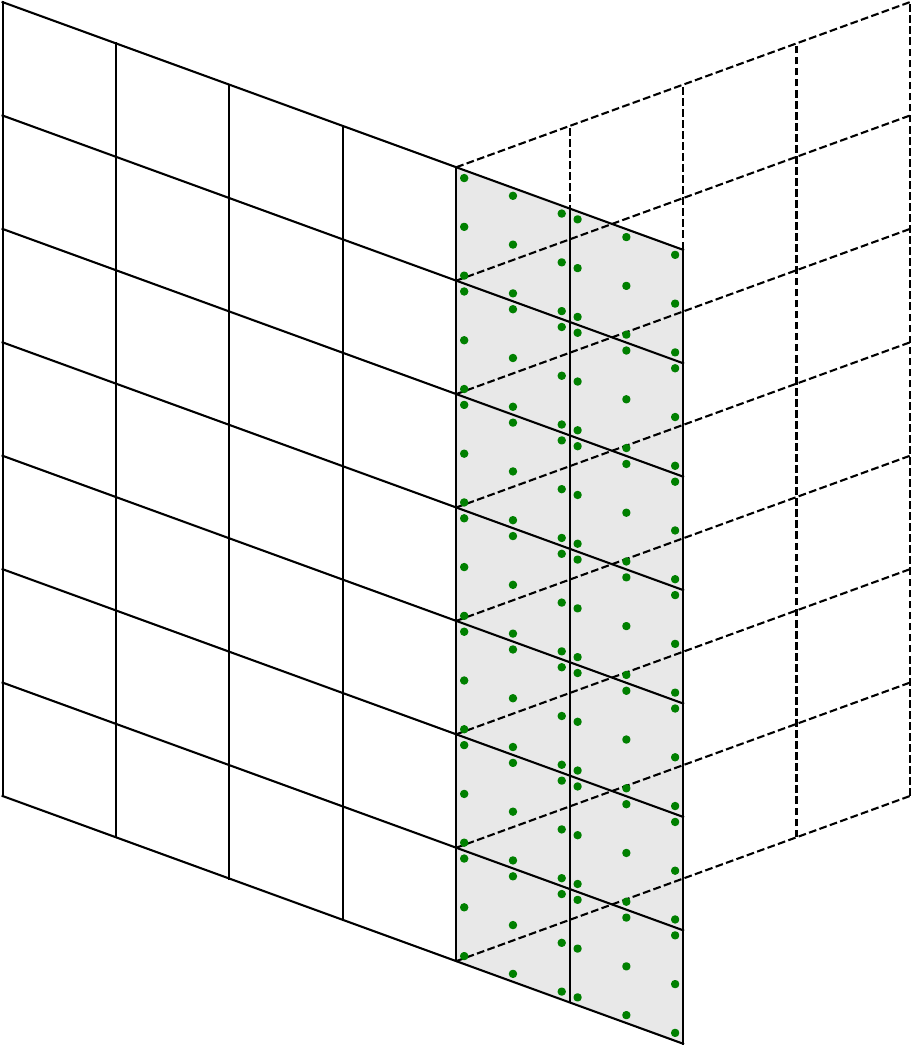}
            \subcaption{User-Defined Quadrature}
            \label{fig:bc-example:quad}
        \end{subfigure}
        \begin{subfigure}[t]{0.8\linewidth}
            \centering
            \includegraphics[width=0.9\textwidth]{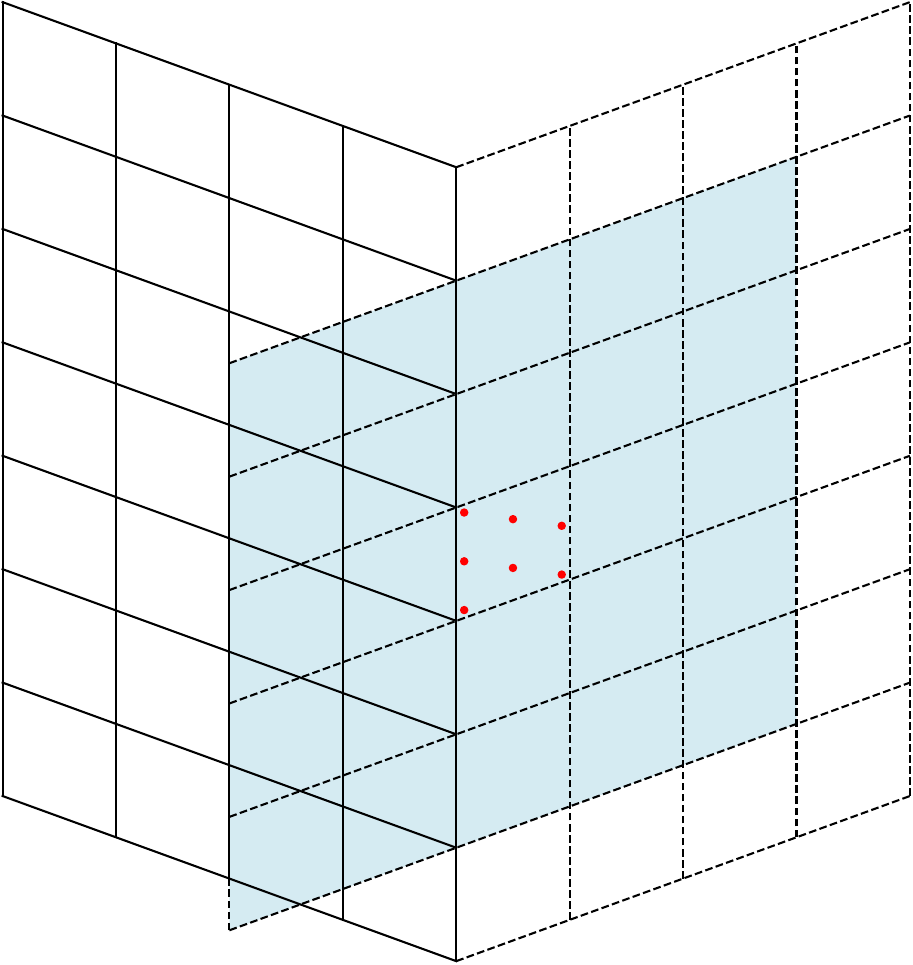}
            \subcaption{User-Defined Interpolation}
            \label{fig:bc-example:interp}
        \end{subfigure}
    \end{subfigure}
    \caption{Common BCs for block-structured grids}
    \label{fig:bc-example}
    
\end{figure}

\Cref{fig:bc-example} demonstrates common boundary algorithms implemented in real applications, ranging from convolutions on images to PDE solvers on structured or block-structured grids. 
The algorithms shown in Figures \ref{fig:bc-example:zero}, \ref{fig:bc-example:halo}, and \ref{fig:bc-example:cs} are widely adopted by popular frameworks and applications even beyond the domain of PDE solver. Meanwhile, the algorithms depicted in Figures \ref{fig:bc-example:pf}, \ref{fig:bc-example:avg}, \ref{fig:bc-example:quad}, and \ref{fig:bc-example:interp} are also commonly utilized in different PDE solvers, but the detailed representation and parameters vary significantly through different governing equations, grids and discretization methods. 
Notably, the degree of freedom for boundary algorithms comes from not only the increase in basic types, but also the flexible combination of all these types. 
For example, users may introduce the circular type to x-direction ghost cells and the padding zero type to y-direction ghost cells when modeling a cylinder geometry. 

\subsection{User-Defined Interpolation Example}
\label{sec:bg:motivated_example}
We take the finite-volume SWE solver HOPE \cite{HOPE} as an example to demonstrate complexity of general user-defined BCs on block-structured grids. 
In order to deal with the discontinuity of different parts of the cubed-sphere grid, for each element in global \textit{ghost} region, users select one or several \textbf{reconstruction points} in the element's corresponding area, and view their numeric quadrature as the value for the whole element. \Cref{fig:bc-example:quad} shows the Gaussian quadrature points that the HOPE method uses for the reconstruction of ghost cells on the left block. 
For calculating the reconstruction points, users view them from other parts of the grid (the right block), translate their position through coordinate transformation between different parts, and carry out another interpolation progress from the data of the right part. 

Despite the difference in dependencies and weights for the interpolation and quadrature procedures, the main complexity of HOPE lies in the use of the ghost cell values in the right part for calculating reconstruction points derived from the left part (shown in \Cref{fig:bc-example:interp}), while other solvers \cite{chen2008mcv, ullrich2010high, GungHo} only use data cells. 
This configuration escalates the difficulty of finding exact solutions to the same level as solving a system of linear equations. 
In practice, HOPE first initializes all of the ghost cells to zero and carries out a fixed number of iterations (10 times) to obtain a fairly approximate solution, which is equivalent to solving the linear system using Jacobi method during each calculation. 
Furthermore, HOPE proves that the whole iterating process is still a linear operator on normal data cells to calculate ghost cells, therefore it calculate the weights in advance and stores them down in Compressed-Sparse-Row (CSR) format, therefore the whole iterating process can be expressed as Sparse Matrix-Vector multiplication (SpMV). We reference these two algorithms as HOPE(I) (Iterative) and HOPE(M) (Matrix) in later description. 

\subsection{Opportunities and Challenges}
In this section, we demonstrate our intuitive ideas for supporting all kinds of boundary conditions. Ideas and their corresponding opportunities and challenges are summarized in four aspects as below: 
\label{sec:bg:oppor}
\subsubsection{Universality} Combination of different BCs are necessary when dealing with real-world complicated geometry. Therefore, an unified representation for all types of boundary algorithms will boost programmer's productivity on processing complicated boundary problems. The sparse matrix is a good abstraction to represent BC and has been used in HOPE-matrix method. However, its versatility among general finite-volume methods remains unexplored. 
\subsubsection{Simplicity} 
As mentioned in \Cref{tab:LOC}, the workload of writing BC code constitutes a significant or even the main portion of the entire programming workload, therefore the simplicity of boundary configuration is vital for productivity. 
A declarative configuration style is desired, where users only designate the boundary regions and their corresponding computation patterns. The distributed implementation \cite{bisbas2025automated} of Devito DSL \cite{devito} meet the requirements, but it only support simple halo-copying. The new Python implementation of FV3 dynamic core \cite{ben2022productive} uses a cubed-sphere specific communication library for the configuration. But when other grids, discretization methods or distribution schemes are introduced, users have to work on the low-level MPI-based library code, which will block the whole development progress. 
\subsubsection{Efficiency} Users may declare a complex boundary region (ring-like ghost region) that share the same computation pattern, therefore it is the compiler's duty to generate code for traversing these regions efficiently. What is more, there exists two specific optimizing opportunities for BC computation
\begin{itemize}
    \item The boundary conditions may consists of simple types which do not need to iterate over a real sparse matrix (pure function, symmetric and circular). Therefore, generate matrix-free code for these BCs can reduce memory cost while improve execution efficiency. 
    \item We notice that BC computations do not appear alone. They are always followed by inner calculations. Additionally, the assignments of BC computations are only used inside its corresponding operator. Therefore, \textbf{non-equivalent} transformations could be taken into account for BC computations as long as they do not affect the inner calculation process. This feature can benefit both the configuration simplicity and the execution efficiency.
\end{itemize}

\subsubsection{Scalability} For a framework or DSL, scalability is determined by both its execution efficiency and its ease of programming at scale. A major challenge lies in designing abstractions that allow developers to scale their applications with only minor additions with the underlying data distribution configurations. Achieving scalability also requires a highly capable communication backend — one that can perform rapid halo exchanges while retaining the flexibility to handle irregular data retrieval for unstructured boundary conditions.

\section{SpMV Representation of Boundary Conditions}
\label{sec:analysis}

In this section, we formalize the representation of different boundary condition algorithms 
and demonstrate their interconnection with sparse matrix-vector multiplication (SpMV). 

\subsection{Definitions}
Shown in~\cite{HOPE, ullrich2010high, chen2008mcv}, and in \Cref{fig:bc-example:quad} and~\Cref{fig:bc-example:interp}, the user-defined boundary condition algorithms compute the values on the extended grid cells of each panel in the cubed-sphere grid. 
Therefore, we structure the \textit{full} region of a  \frameworkname{}'s grid variable $y$ into 2 distinct regions: \textit{data} and \textit{ghost}. 
We denote the coordinate sets of \textit{data} region as $R^{y}_D$ and the coordinate sets of \textit{ghost} region as $R^{y}_G$. Then we can use a vector of the size $|R^{y}_F|$ to represent the output value of a user-defined boundary condition algorithm, where $R^{y}_F = R^{y}_D \cup R^{y}_G$. 

When in the context of distributed memory parallelism (DMP), each process owns a sub-part of the whole \textit{data} region, denoted as $R^{y}_{d_p}$ for process $p$ and variable $y$. We have the relationship that $\cup_{p} R^y_{d_p}= R^{y}_{D}$. The process $p$ also maintains the data in the local \textit{halo} region $R^{y}_{h_p}$ for later calculation, and $\cup_{p} (R^y_{f_p}) = R^y_F$, where $R^y_{f_p} = R^y_{d_p} \cup R^y_{h_p}$. 
Using all these definitions, we can formulate a distributed user-defined boundary condition algorithms as computing $$y_{(|R^{y}_{f_p}|)} = F(x_1, x_2, ...)$$ in each process, where the input parameter $x_i$ has a global data region of ${R^{x_i}_{D}}$. 

\subsection{Representation}
\label{sec:spmv-rep}
A common and powerful paradigm of computing $y$ using $x$ is matrix-vector multiplication, and we list below how to implement all kinds of common boundary conditions using this paradigm: 
$$ F(x)=A_{(|R^y_{f_p}|,|R^x_D|)}\times x_{(|R^x_D|)}+b_{(|R^y_{f_p}|)}$$

\begin{itemize}
    \item \textbf{Padding with Zeroes} in \Cref{fig:bc-example:zero} is the simplest boundary condition for the sub-part of the \textit{ghost} region that process $p$ locally holds. 
    
    For index $i \in R^y_{h_p} \cap R^y_{G}$ and $j \in R^x_D$, $A_{ij} = b_i = 0$
    
    \item \textbf{Halo Copying} in \Cref{fig:bc-example:halo} is always introduced by distributed memory parallelism and is fully researched and optimized by manual implementations and dsl implementations. In this case, $R^y = R^x$. 
    
    For index $i \in R_{h_p} \cap R_{D}$ and $j \in R_D$,  $A_{ij} = \delta_{ij},\ b_i = 0$. 
    
    \item \textbf{Circular and Symmetric Boundaries} in \Cref{fig:bc-example:cs} also demand $R^y = R^x$. 
    For index $i \in R_{h_p} \cap R_{G}$ and $j \in R_D$, 
    $$A_{ij} = Boolean(j = fmap(i)),\ b_i = 0$$
    where $fmap(i)$ is a simple index mapping. When in the one-dimensional case where $R_D = \{0, 1, ..., |R_D|-1\}$, $fmap$ for the left ghost region has the form $$fmap(-x)=\begin{cases}x-1& \text{symmetric}\\ 
        |R_D|-x& \text{circular}
    \end{cases}, 0 < x < |R_D|$$
    
    \item The \textbf{Pure Function Boundary} in \Cref{fig:bc-example:pf} is common in testing the performance of PDE solvers on equations with analytical solution. For example, building a boundary with Dirichlet Conditions. 
    
    For index $i \in R^y_{h_p} \cap R^y_{G}$ and $j \in R^x_D$,  $A_{ij} = 0,\ b_i = udf(i)$
    
    \item \textbf{Block Edge Averaging} in \Cref{fig:bc-example:avg} occurs when there are several grid cells located on different parts of the block-structured grid that share the same physical location~\cite{chen2008mcv}, therefore the condition $R^y = R^x$ holds. 
    We pick all these specific cells on the block edges to form a set $R_{SD}$, Therefore the relationships between these cells can be viewed as a sub-matrix $Sync_{|R_{SD}|, |R_{SD}|}$. $Sync$ has a great sparsity. Normally each row of the $Sync$ has only 2 or 3 non-zero elements. 
    For index $i \in R_{d_p}$ and $j \in R_D$, 
    $$A_{ij} = \begin{cases}
        Sync_{ij}&  j \in R_{SD} \\
        0, &  j \notin R_{SD}\\
    \end{cases}, \ b_i = 0$$

    \item The \textbf{User Defined Interpolation Boundary} in \Cref{fig:bc-example:quad} and \Cref{fig:bc-example:interp} appears in realistic applications of PDE solvers on block-structured grids~\cite{HOPE, chen2008mcv, ullrich2010high}. 
    The authors of HOPE~\cite{HOPE} prove in their article's Appendix that their boundary algorithm is equivalent to multiplying a matrix of shape ($|R_G|, |R_D|$) on variable $x$ to get ghost points on $y$, where $R^y = R^x$. They reach this conclusion by defining the set of reconstruction points as $Pts$ and figuring out that their Tensor product polynomial (TPP) reconstruction scheme is equivalent as a linear mapping $Recons_{|Pts|, |R_D|}$ for calculating reconstruction points, and their Gaussian quadrature integration scheme can be viewed as another linear mapping $Agg_{|R_G|, |Pts|}$ for calculating ghost cell values. Multiplying $Agg$ and $Recons$ together, they obtain the final linear mapping as the matrix $Itp_{|R_G|, |R_D|}$. 

    We find out that the simplification process of HOPE is much more general than the method itself. As the set $Pts$ does not affect the final matrix $Itp$, we claim that the different methods, order of accuracy and reconstruction points $Pts$ only affect the detailed value of $Itp$. 
    Therefore, by asking users to pre-process the matrix $Itp$,
    for index $i \in R_{h_p} \cap R_{D}$ and $j \in R_D$, \frameworkname{} represents the boundary as 
    $$A_{ij} = Itp_{ij},\ b_i = 0$$

\end{itemize}
In summary, we illustrate six types of common boundary conditions in block-structured grids and show their implementation with the abstraction of SpMV. 
Notably, the construction of these matrices heavily relies on conditional branching and case-by-case analysis, and the final result of matrix A is of great sparsity. This motivates our design of the \textit{basic sub-matrix} abstraction which is described in detail in \Cref{sec:design}.

\section{\frameworkname{} Design}
\label{sec:design}
Based on the observation mentioned above, we propose \frameworkname{}, our domain-specific language and its corresponding compiler and runtime library for solving boundary problems on distributed block-structured girds. 

\begin{figure}[htbp]
    \centering
        \includegraphics[width=0.9\linewidth]{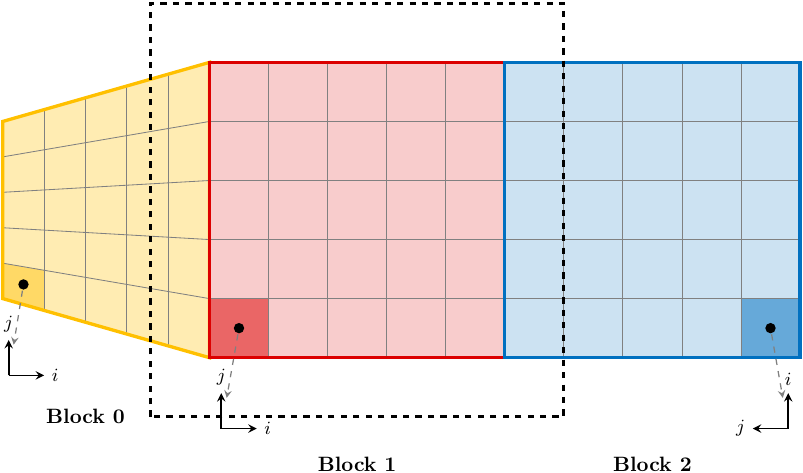}
        \caption{Example Block-Structured Grid}
        \label{fig:design-example}
\end{figure}

We demonstrate our design using the 3 blocked 2-D Cartesian grid example shown in \Cref{fig:design-example}. 
In contrast with block 1, block 0 has a similar axis order but its shape is trapezoidal, therefore the left \textit{ghost} region of block 1 need an user-defined interpolation BC. The geometry of block 2 is the same with block 1, but its axes are rotated, therefore the right \textit{ghost} region BC of block 1 is in the form of a simple mapping. The top and bottom \textit{ghost} region BC is set to zero. The \textit{data} region size is set to 5 $\times$ 5 and the \textit{full} region size is set to 7 $\times$ 7. 

\subsection{Modular Boundary Matrix Construction}
\subsubsection{Region Representation}
\label{sec:design:rep}
Before demonstrating our interface for configuring matrix, we first illustrate our representation for iteration space and structural grid cells. 

\frameworkname{} will first suppose the existence of a multi-dimensional Cartesian grid which covers the range of $[Start_{d}, Stop_{d})$ in each dimension $d$. 
Then, \frameworkname{} configure the \textit{full} region and the \textit{data} region of each structural part of a block-structured grid as two slices of the global grid, and the \textit{ghost} region is the difference between \textit{full} and \textit{data}. The slice is capable of not only decreasing the area size, but also applying bigger strides on each dimension to get a coarser grid. We use ($R\_start_d, R\_stop_d, R\_step_d$) for describing the slicing in each dimension for the grid. 

While the $start$ and $stop$ variables are widely used for grid description~\cite{falgout2002hypre, devito}, the $step$ variable is innovative in PDE solver frameworks and dsls. We list two benefits for the introduction of $step$:

\begin{itemize}
    \item Naturally support grid staggering technique with \textit{step} = 2. Taking 2-D problems as examples, in our language, cell-centered, edge-centered, and nodal variables of structural grids now have nothing different from each other except for the detailed shape information. This feature is essential for the development of Finite Volume Method (FVM) solvers of PDE. 
    \item Naturally support multi-grid method with \textit{step} = $2^k$.
\end{itemize}

\subsubsection{Basic Data Types and Operations}
\label{sec:design:dtype}
\frameworkname supports a wide range of statements in the procedures of calculating each element of $y$ for $y=A \times x+b$. 
These statements include creating temporary variables, if statements, while-loop statements without break or continue, and assign statements for temporary variables. 
We use $E_I$ and $E_F$ for describing the expression data type of the integer and floating parameters in the above statements. 

The opportunities for extra optimization in the compile-time arise
when we find a subset of the expressions which only rely on the coordinate indices of the current $y$ elements and other integer parameters that remain constant over the whole SpMV procedure ($R\_start_d, R\_stop_d, R\_step_d$). 
We call these expressions as compile-time resolvable expressions and define two new sub-types $C_I$ and $C_B$ for Integer and Boolean respectively. The supported operations of the sub-types are listed below using BNF description. Note that the expressions loaded from temporary variables are not treated as resolvable. 
\begin{align*}
C_B \Coloneqq &   && C_I < C_I \mid C_I = C_I  \\
							& | && C_B \land C_B \mid C_B \lor C_B \mid \lnot C_B \\
C_I   \Coloneqq &   && C_I \times \text{Literal} \mid C_I \div \text{Literal} \mid C_I \text{ mod } \text{Literal} \\
							& | && C_I + C_I \mid \mathbf{max}(C_I, C_I) \mid \mathbf{min}(C_I, C_I) \\
							& | && \text{Name} \mid \text{Literal}
\end{align*}
\subsubsection{Basic Sub-Matrix/Vector API}
As directly using the statements of data types $E_I$, $E_F$, $C_I$, $C_B$ to implement a whole SpMV for \Cref{fig:design-example} is still tedious for programmers, \textit{basic sub-matrix/vector} are introduced to decouple the boundary conditions into several sub-problems which are simple enough to omit the detailed SpMV procedure. 

\begin{table}[t]
    \centering
    \begin{tblr}{colspec={Q[c,m] Q[c,m] Q[c,m]}}
        \toprule
         Interface & Parameter Type & Description \\
         \midrule
         {\texttt{is\_valid}} & List[$C_I$] $\rightarrow$ $C_B$ & Is in coverage area ($R_{g_p} \cap R_D$, e.g.) \\
         {\texttt{iterator}} & List[$C_I$] $\rightarrow$ Iter &  A sparse row of matrix ($\delta_{ij}, Itp_{ij}$, e.g.) \\ 
         {\texttt{n\_group}} & $\rightarrow$ $E_I$ & Property for distributed communication \\
         \bottomrule
    \end{tblr}
    \caption{Basic Sub-matrix APIs}
    \label{tab:basic_api}
\end{table}

\Cref{tab:basic_api} shows the core apis of a \textit{basic sub-matrix}. 
Comparing with a general meaning row-iterator of sparse matrix, the \textit{basic sub-matrix} needs an additional Boolean expression to describe its responsible area in the whole matrix, therefore the return of \texttt{iterator} interface can be extremely simple. 
We will describe the usage of \texttt{n\_group} in \Cref{sec:impl:library}. 

The \textit{basic sub-vector} api is similar with \textit{basic sub-matrix}. Instead of \texttt{iterator} and \texttt{n\_group} interface, \textit{basic sub-vector} only requires user to designate an expression with the indices as input. 

\subsubsection{Row Iterator API}
\label{sec:design:interface}

As for row iterator, two formats are supported as the return value of the \textit{basic sub-matrix} api for describing structured and unstructured part of boundary conditions. 
\begin{enumerate}
    \item Users can directly return a compile-time static list of non-zero elements as result, which can cover simple BCs like \textbf{Circular and Symmetric Boundaries}. This interface is called as \texttt{F\_LIST}. 
    \item For the unstructured and user-defined boundary condition patterns, we demonstrate four interfaces, \texttt{F\_NNZ}, \texttt{F\_COL} and \texttt{F\_DATA} and \texttt{F\_GID} to describe an iterator over non-zero elements.
\end{enumerate}

 \Cref{tab:func_api} shows the meanings of the five function interface. As we can construct the row iterator either in compile time or in real execution using the provided function interface, we have the default implementation of SpMV and its detailed procedure can be omitted by user. 

\begin{table}[t]
    \centering
    \begin{tblr}{colspec={Q[c,m] Q[l,m] Q[c,m]}}
        \toprule
         Interface & Parameter  & Description \\
         \midrule
         {\texttt{F\_LIST}} & {\text{indices}: List[$C_I$] $\rightarrow$ \\ elements: List[$T$]} & {return a static \\ number of entries} \\ 
         {$T$} & Tuple[List[$C_I$], $E_F$, $E_I$] & {column, data and gid}\\
         \midrule
         {\texttt{F\_NNZ}} & {\text{indices}: List[$C_I$] $\rightarrow$ \\ \text{nnz}: $E_I$, \\ \text{info}: Tuple[$E_I$]} & {return the length of \\ the dynamic iterator} \\ 
         \midrule
         {\texttt{F\_COL}} & {\text{indices}: List[$C_I$], \\ nnz: $E_I$, \\ info: Tuple[$E_I$], \\ offset: $E_I$ $\rightarrow$ \\ \text{col}: List[$E_I$]} & {the column of \\ the offset-th entry \\ of the iterator} \\ 
         {\texttt{F\_DATA}} & {List[$C_I$], $E_I$, Tuple[$E_I$], $E_I$ \\ $\rightarrow$ data: $E_F$} & {entry property of data} \\ 
         {\texttt{F\_GID}} & {List[$C_I$], $E_I$, Tuple[$E_I$], $E_I$ \\ $\rightarrow$ gid: $E_I$} & {entry property for \\ distributed communication} \\ 
         \bottomrule
    \end{tblr}
    \caption{Parameter and Description of the functions for row iterator building}
    \label{tab:func_api}
\end{table}

\subsubsection{Example BC representation}
Algorithm~\ref{alg:mat_impl} shows the pseudo-code implementation of the three boundary conditions in \Cref{fig:design-example}. The \texttt{n\_group} and \texttt{F\_GID} interfaces are omitted for simplicity. 
Notably, our unstructured api can represent generalized \textbf{row-based} sparse matrices as the user implement interfaces for the detailed storage format, and we provide the basic implementation for CSR and ELL format in advance, therefore users can configure the boundary using one line like \texttt{BoundaryCSR(ptr\_arr, col\_arr, data\_arr, calc\_addr, extract)}. 

\algtext*{EndIf}  
\algtext*{EndFor} 
\algtext*{EndWhile} 
\algtext*{EndFunction}
\newcommand*\Let[2]{\State #1 $\gets$ #2}
\begin{algorithm}[h]
    \caption{Pseudo-code for the implementation of \textit{basic sub-matrix} to describe boundaries in \Cref{fig:design-example}}
    \label{alg:mat_impl}
    \footnotesize
    \begin{algorithmic}[1]
        \Let{$R_D$}{$(0, 3, 1) \times(0, 5, 1)\times(0, 5, 1)$}     \Comment{3D grids are used}
        \Let{$R^i_D$}{$(i, i+1, 1) \times(0, 5, 1)\times(0, 5, 1)$}  \Comment{for blocks of 2d grids}
        \Let{$R^i_F$}{$(i, i+1, 1) \times(-1, 6, 1)\times(-1, 6, 1)$}
        \Let{$R_{right}^1$}{$(1, 2, 1) \times(0, 5, 1)\times(5, 6, 1)$}
        \Let{$R_{left}^1$}{$(1, 2, 1) \times(0, 5, 1)\times(-1, 0, 1)$}
        \State
        \Function {BC\_Right.is\_valid}{\text{indices}: \text{List[$C_I$]}}
        \State \Return{\textsc{In\_Area}(\text{indices}, $R_{right}^1$)} 
        \EndFunction
        \Function {BC\_Left.is\_valid}{\text{indices}: \text{List[$C_I$]}}
        \State \Return{\textsc{In\_Area}(\text{indices}, $R_{left}^1$)} 
        \EndFunction
        \Function {BC\_Zero.is\_valid}{\text{indices}: \text{List[$C_I$]}}
        \State \Return{\textsc{In\_Area}(\text{indices}, $R_F^1$) $\land$}
        \State $\lnot${\textsc{In\_Area}(\text{indices}, $R_D^1$)} $\land$
        \State $\lnot${\textsc{In\_Area}(\text{indices}, $R_{right}^1$)} $\land$
        \State $\lnot${\textsc{In\_Area}(\text{indices}, $R_{left}^1$)} 
        \EndFunction
        \State
        \Function {BC\_Right.iterator}{\text{indices}: \text{List[$C_I$]}}
        \Function {f\_list}{\text{indices}}
        \State \Return{[([2, 9-\text{indices}[2], \text{indices}[1]], 1)]} \Comment{indices mapping}
        \EndFunction
        \State \Return{wrap\_to\_iterator(\textsc{f\_list})}
        \EndFunction
        \State
        \Let {ptr\_arr}{\text{Array[n\_row+1]}} \Comment{use CSR matrix to }
        \Let {col\_arr}{\text{Array[nnz]}}   \Comment{represent left boundary}
        \Let {data\_arr}{\text{Array[nnz]}}
        \Function {BC\_Left.iterator}{\text{indices}: \text{List[$C_I$]}}
        \Function {f\_nnz}{\text{indices}}
        \Let{pt}{calc\_addr(indices, n\_row)}
        \Let{addr\_1}{row\_ptr\_arr[pt]}
        \Let{addr\_2}{row\_ptr\_arr[pt+1]}
        \Let{row\_nnz}{addr\_2 - addr\_1}
        \State \Return{row\_nnz, addr\_1}
        \EndFunction
        \Function {f\_col}{\text{indices}, row\_nnz, info, offset}
        \Let{addr\_1}{info}
        \Let{raw\_col}{col\_arr[addr\_1+offset]} \Comment{type is $E_I$}
        \Let{col} {extract(raw\_col)} \Comment{type is List[$E_I$]}
        \State \Return{col}
        \EndFunction
        
        \Function {f\_data}{\text{indices}, row\_nnz, info, offset}
        \Let{addr\_1}{info}
        \State \Return{data\_arr[addr\_1+offset]}
        \EndFunction
        \State \Return{wrap\_to\_iterator(\textsc{f\_nnz}, \textsc{f\_col}, \textsc{f\_data})}
        \EndFunction
        \Function {BC\_Zero.iterator}{\text{indices}: \text{List[$C_I$]}}
        \Function {f\_list}{\text{indices}}
        \State \Return{[]}
        \EndFunction
        \State \Return{wrap\_to\_iterator(\textsc{f\_list})}
        \EndFunction
    \end{algorithmic}
    
\end{algorithm}

\subsection{SpMV Code Generation}
\label{sec:design-codegen}
As mentioned in \Cref{sec:bg:oppor}, optimization opportunities arise when BCs have structured sub-parts and can be hard-coded into the program. We use the multi-stage programming techniques here to generate matrix-free code for these BCs. 

Multi-stage programming (commonly known as staging) has emerged as a promising technique for developing embedded domain-specific languages. In this paradigm, programs are typically divided into two distinct phases: 
\begin{itemize}
    \item Stage 1 (compile-time): A partial evaluation of the user code occurs and produces an optimized intermediate program.
    \item Stage 2 (runtime): The generated intermediate program is executed to compute the final results.
\end{itemize}

The SpMV code generation occurs in Stage 1, when our compiler can easily distinguish between the \texttt{iterator} types and generate different code for them. 
Furthermore, we can discuss the relationships of these \textbf{overlapping} expressions in $originSet$ with each other and get a new \textbf{non-overlapping} expression set $newSet$, whose additional features are listed below: 
\begin{equation}
\begin{split}
    \lor_{newSet} e_i \equiv \textsc{True}
\end{split}
\end{equation}
\begin{equation}
\begin{split}
\forall & e_i \in newSet, \forall e_j \in originSet \\
&\text{either }e_i \rightarrow e_j \equiv \textsc{True} \text{ or } e_i \rightarrow \lnot e_j \equiv \textsc{True}
\end{split}
\end{equation}

By iterating over the power set of the given \textit{basic sub-matrices/sub-vectors}, we can easily build the $newSet$ by discussing whether the basic sub-matrix/sub-vector belongs to current subsets of the complete collections.  
Then non-overlapping expressions are formed and lifted to the beginning of the loop using a huge if-elsif-else statement, and multiple final assignments points to the vector $y$ are generated corresponding with these expressions. This transformation gathers different BC computations together in certain cell areas and provides more opportunities for later ordinary optimizations and specific non-equivalent optimizations in \Cref{sec:global-opt}. 

In real implementations, we find that the size of power set grows exponentially as the number of \textit{basic sub-matrices/sub-vectors} increases, and most of the derived Boolean expressions can be proved to be \textbf{False} during the stage-1 compilation, since in real configurations the valid areas of different sub-matrices are merely intersected. Therefore, we utilize the \textit{backtrack and discuss} algorithm shown in the \plainmtos DSL \cite{cao2023mat2stencil} to reduce the compiling cost and the final number of branches, since this algorithm can early-exit when a sub-part of a Boolean expression can be proved to be false by Z3 SMT solver \cite{Z3} and skip generating code for their branches. 
Therefore we can directly obtain a SpMV for-loop with only a handful number of branches inside. 

\subsection{Optimization for Boundary Condition Computation \label{sec:global-opt} }
In this section, we introduce our techniques to optimize the generated code of boundary condition in the runtime stage. 
Our key idea is to discover optimization opportunities from a global view of the whole program, then we can rely on less instructions from users and gain flexibility in programming. 

\begin{figure}[htbp]
\centering
    \begin{subfigure}[b]{0.48\linewidth}
    \centering
        \includegraphics[width=\textwidth]{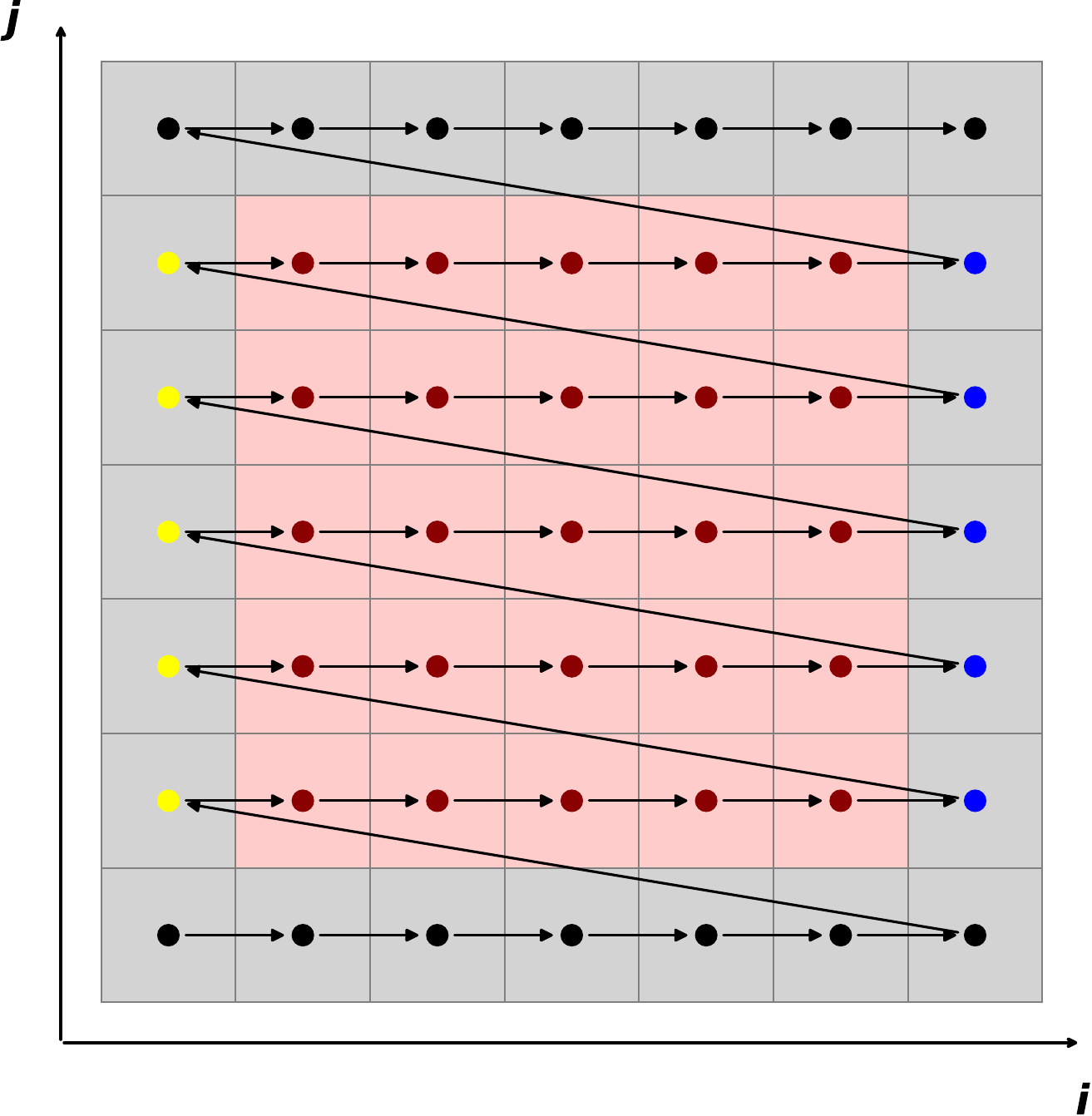}
        \caption{Origin}
        \label{fig:design-opt-1}
    \end{subfigure}
    \hfill
    \begin{subfigure}[b]{0.48\linewidth}
    \centering
        \includegraphics[width=\textwidth]{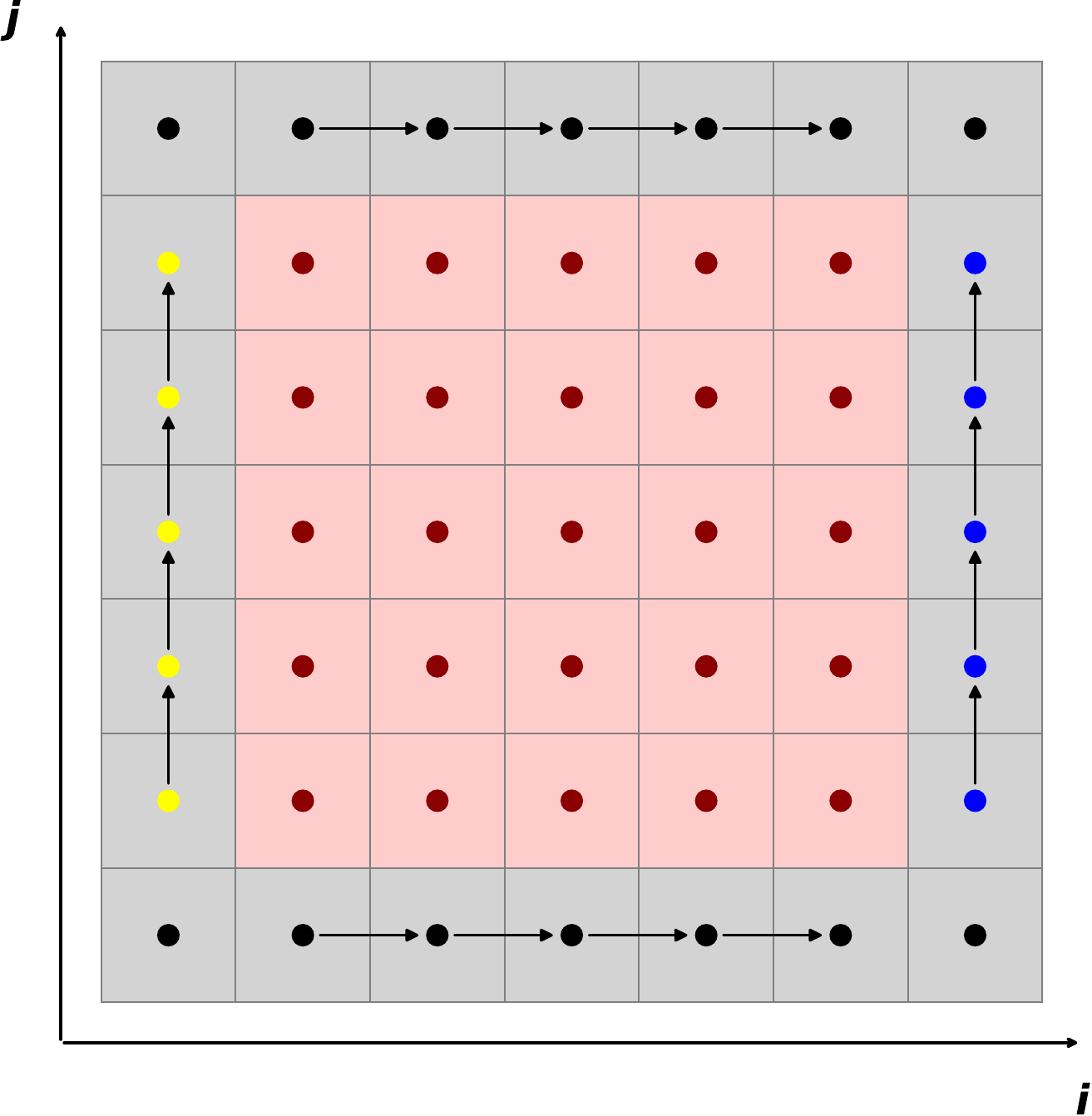}
        \caption{After}
        \label{fig:design-opt-2}
    \end{subfigure}
\caption{Example on simplification of the iteration space of boundary condition algorithm in \Cref{fig:design-example}}
\label{fig:global_opt}
\end{figure}

The left part of Figure~\ref{fig:global_opt} shows the iteration space of the generated boundary condition code. 
The initial iteration space has redundant visits to the empty part of the sparse matrix $A$ (central red parts) and parts of the \textit{ghost} regions that may never be used by later calculations (corner gray parts for 2d-5 points stencil). 

Expressing boundary and inner computations in a unified IR enables us to extract the program's complete dependency graph using polyhedral analysis and Presburger arithmetic. In the optimization phase, we capture fine-grained read-after-write (RAW) dependencies from boundary assignments ($S_1$) to inner calculations ($S_2$), formulated as:
\begin{equation}
\begin{split}
    D(S_{1}, S_2) = \{(i_1, i_2)|S_{2} \text{ at iteration }i_2\\ \text{ depends on } S_{1}\text{ at iteration }i_{1} \}
\end{split}
\end{equation}
where $i_k$ is the iteration vector of statement $S_k$ in loop $L_k$, with $L_2$ following $L_1$. For each boundary write $W_i$, we union the dependencies of all subsequent dependent reads $R_j$ to construct its effective iteration space:
\begin{equation}
    U(W_{i}) = \bigcup_{R_{j}}\left\{i|\exists j, (i, j)\in D(W_{i}, R_{j})\right\}
\end{equation}
This effective space $U(W_i)$ allows us to safely eliminate unnecessary boundary assignments (gray regions in \Cref{fig:global_opt}). We then partition the remaining space $U(W_i)$ into disjoint convex sets. Because calculations within these separate subsets are embarrassingly parallel, their code can be independently instantiated. As shown in \Cref{fig:global_opt} (right), all the user-defined interpolation, user-defined simple mapping and padding-zero boundaries are neatly split into simple Cartesian sub-spaces.

We emphasize that \frameworkname{} compiler automatically removes redundancy of the iteration space of boundary condition and generates high-efficient codes of iterating that are comparable with manual implementation. We use FreeTensor~\cite{tang2022freetensor} for representing BC code and inner calculation code together and for carrying out polyhedral analysis.  

\section{Implementation}
\label{sec:impl}
To seamlessly deploy \frameworkname-generated solvers onto large-scale clusters, we bridge the gap between high-level matrix representations and low-level distributed infrastructure through several implementation strategies. 

\subsection{Representation Extension for Real-World Workloads}
Real-world PDE solvers require abstractions beyond simple scalar grids. 
We extend \frameworkname's internal representations to natively support multi-dimensional arrays at each grid point, allowing the compiler to co-optimize tightly coupled physical quantities (e.g., multi-component velocity vectors). 
Furthermore, we map the global iteration spaces to distributed memory layouts by injecting partition-aware coordinates. This spatial mapping enables the \frameworkname DSL to isolate standard halo-exchange scenarios from complex, unstructured boundary interpolations at the matrix representation level, significantly reducing the programming burden on users.

\subsection{Domain-Specific Communication Backend}
\label{sec:impl:library}
Evaluating the SpMV abstraction $y=Ax+b$ in a distributed environment necessitates fetching remote data entries for vector $x$. Instead of relying on general-purpose sparse matrix communication routines—which incur heavy metadata and packing overheads (e.g., using hash tables to deduplicate fetch indices)—we exploit the domain-specific properties of PDE solvers.
Leveraging the polyhedral analysis from \Cref{sec:global-opt}, which partitions boundary operations into convex subsets, we guarantee that the remote data dependencies for each subset remain strictly compact in physical space. \frameworkname automatically intercepts these remote dependencies and generates an initialization phase that constructs localized bounding boxes around the required remote indices, with the help of \texttt{n\_group} and \texttt{F\_GID} interfaces implemented by the users. During the execution phase, these bounding boxes drive highly efficient, contiguous buffer packing and unpacking. This specialized communicator yields high-bandwidth bulk transmission while bypassing the expensive irregular indexing typically associated with unstructured BCs.

\subsection{Parameterized IR and JIT Specialization}
A major drawback of fine-grained polyhedral analysis is its prohibitive compilation time, particularly when dealing with complex inner calculations. 
To achieve a scalable compilation pipeline, \frameworkname employs an \textit{IR-reusing} technique. 
During the heavy Ahead-of-Time (AOT) compilation stage, problem sizes and partition topological parameters are treated as symbolic variables, producing a generalized, dependency-resolved IR. 
During the initialization phase of execution, a lightweight Just-In-Time (JIT) pass substitutes these symbols with actual runtime dimensions and applies fast, localized optimizations such as constant folding. 
This hybrid approach preserves all advanced polyhedral transformations while amortizing the heavy compilation overhead across cluster nodes.

\section{Evaluation}
\label{sec:eval}
\subsection{Experimental Setup}
\paragraph{Workloads} We evaluate our language, compiler, and runtime with the applications on two grids: cubed-sphere grid, 3-D Cartesian grid with multi-grid solvers. 
For cubed-sphere grids we choose HOPE~\cite{HOPE} and MCV~\cite{chen2008mcv} methods of SWE equation solvers. The order of accuracy configuration for HOPE and MCV methods is chosen as 5 and 3 relatively. 
For multi-grid solvers we choose High Performance Conjugate Gradients (HPCG)~\cite{dongarra2016HPCG}. 

We use the original implementations for the HOPE evaluation \cite{HOPE-soft}, which contains FORTRAN implementations for HOPE-iter and HOPE-matrix algorithm and an optimized Pytorch implementation for HOPE-matrix algorithm. We use a hand-written matrix-free HPCG from \cite{hpcg-impl} for the HPCG evaluation. 
For the MCV method, we use an open-source FORTRAN implementation as the benchmark \cite{MCV_FORTRAN}.

\paragraph{Platform} We evaluate \frameworkname{} on a 24-node cluster, each with 2 Intel Xeon 6258R 28-core CPUs, and 192 GB of main memory. The nodes are interconnected with each other by 100 Gbps InfiniBand HDR. We use GCC 13.2.0 to compile the programs, using g++ for C++ and gfortran for FORTRAN, including our generated codes and baseline implementations. We use pip to install PyTorch with version 2.6.0, and carry out optimizations of the PyTorch code using the interface \texttt{torch.compile}. 

\subsection{Code Length and Development Effort}
We evaluate the programmability of \frameworkname{} by comparing the LOC against manual FORTRAN baselines (Table~\ref{tab:LOC-eva}). 
For boundary condition implementations on the cubed-sphere grid, \frameworkname{} utilizes only 18.7\% to 32.0\% of the FORTRAN LOC across different operators, achieving a geometric average of 23.5\%. 
Coupled with the \plainmtos{} DSL \cite{cao2023mat2stencil} for inner calculations, the end-to-end implementations of the two PDE solvers require merely 36\% of the total kernel LOC. 
These results confirm that integrating \frameworkname{} with existing structured-grid DSLs substantially minimizes the development effort for complex solvers on block-structured grids.
\begin{table}[htbp]
    \centering
    \resizebox{\linewidth}{!}{
    \begin{tblr}{colspec={|Q[c,m]|Q[c,m]|Q[c,m]|Q[c,m]|}, hlines}
         Solver Method & Operator & {Boundary LOCs} & {Calculation LOCs} \\
         \SetCell[r=3]{c} HOPE(I) \cite{HOPE}      & Reconstruction   & 385 vs. 77 &  39 vs. 18 \\
                                                   & Border Flux      & 178 vs. 57 & 159 vs. 42 \\
                                                   & Other            & -          & 542 vs. (290+) 68 \\
         \SetCell[r=3]{c} MCV \cite{chen2008mcv}   & Flux Derivative  & 203 vs. 38 &  49 vs. 67 \\ 
                                                   & Result Averaging & 168 vs. 43 &  23 vs. 10 \\
                                                   & Other            & -          & 317 vs. (290+) 35 \\ 
    \end{tblr}
    }
    \caption{Comparison of lines of code (LOC) for implementing two PDE solvers on cubed-sphere grids using native FORTRAN and our DSLs (\frameworkname{} and \plainmtos{}). The notation ``(290+)'' indicates 290 shared LOC for common infrastructure between the two methods.}
    \label{tab:LOC-eva}
\end{table}
\subsection{Benchmarking}
We use the resolution of $6 \times 90 \times 90$ for SWE solvers and $224^3$ for HPCG application for small-scale benchmarks on a single node. 
\subsubsection{BC computation Comparison}
\begin{figure}[h]
\centering
    \includegraphics[width=0.9\linewidth]{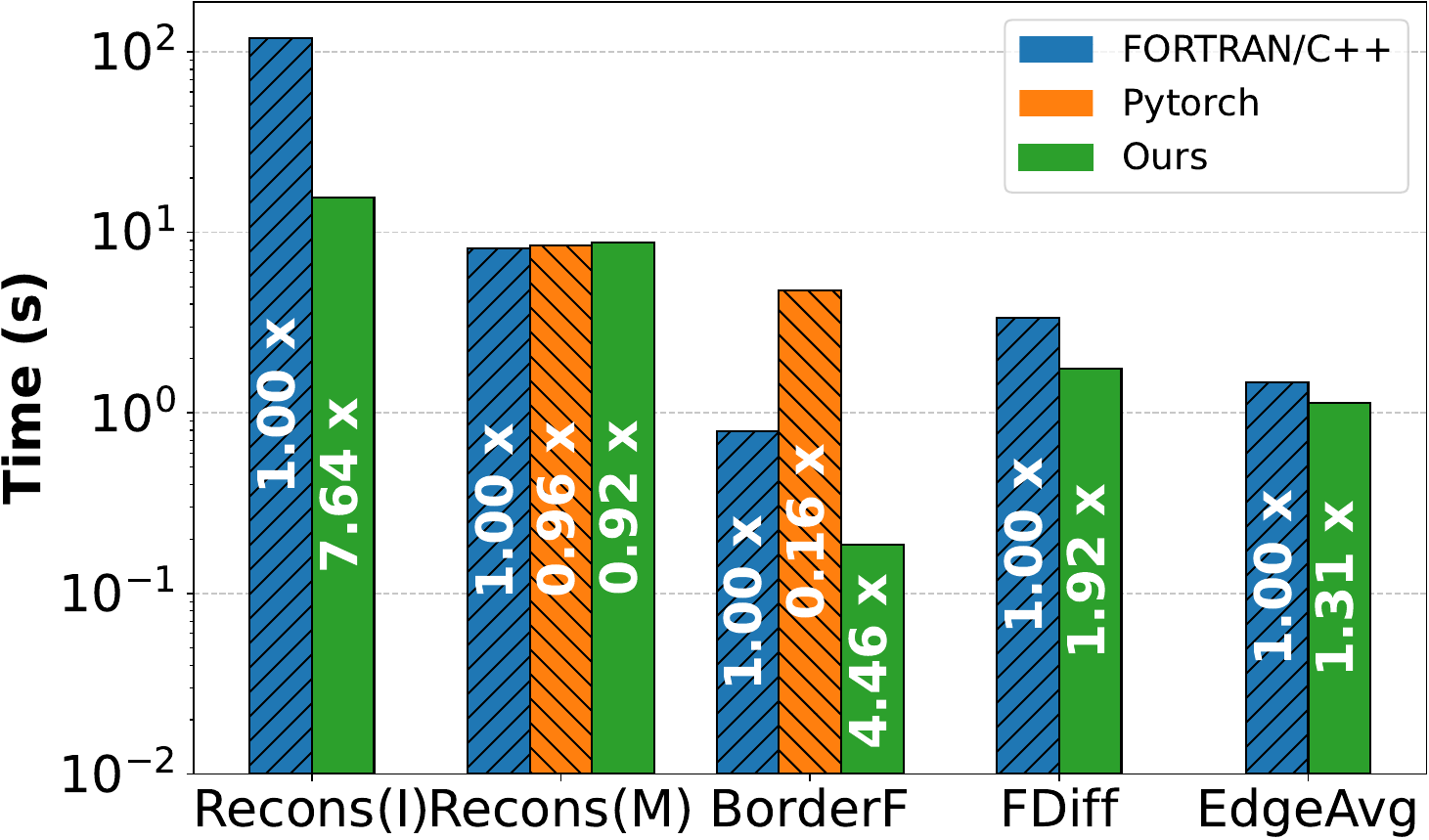}
\caption{BC computation time on a single core. \textbf{Lower is better}. }
\label{fig:eval-bc-single}
\end{figure}

To evaluate the boundary condition (BC) computation cost, we execute 432 time steps for the SWE solvers on a single core and profile the BC computation times, as depicted in \Cref{fig:eval-bc-single}.
The worst-case scenario for \frameworkname{} occurs during the reconstruction operator of the HOPE-matrix algorithm. In this specific case, the BC workload is overwhelmingly dominated by a large Sparse Matrix-Vector Multiplication (SpMV) in CSR format. Although \frameworkname{} incurs a marginal 8\% performance overhead in this highly specific SpMV-bound case, it achieves substantial and consistent speedups across the remaining four types of BC computations, demonstrating a highly favorable overall performance profile.
\subsubsection{End-to-End Comparison}
\begin{figure}[h]
\centering
    \begin{subfigure}[t]{0.9\linewidth}
    \centering
    \includegraphics[width=\textwidth]{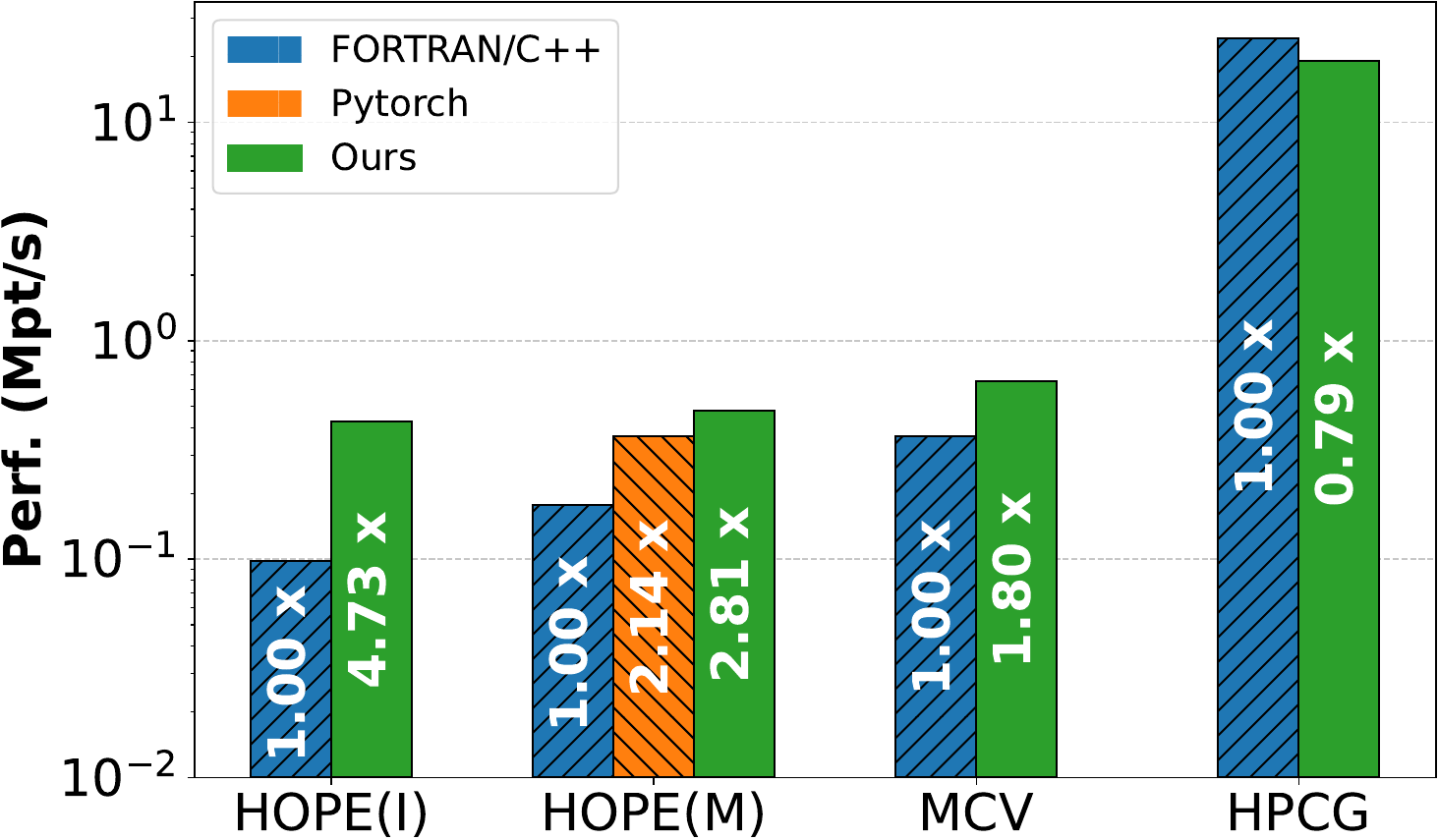}
    \caption{single thread with 1 core}
    \label{fig:eval-single_th}
    \end{subfigure}
    \hfill
    \begin{subfigure}[t]{0.9\linewidth}
    \centering
    \includegraphics[width=\linewidth]{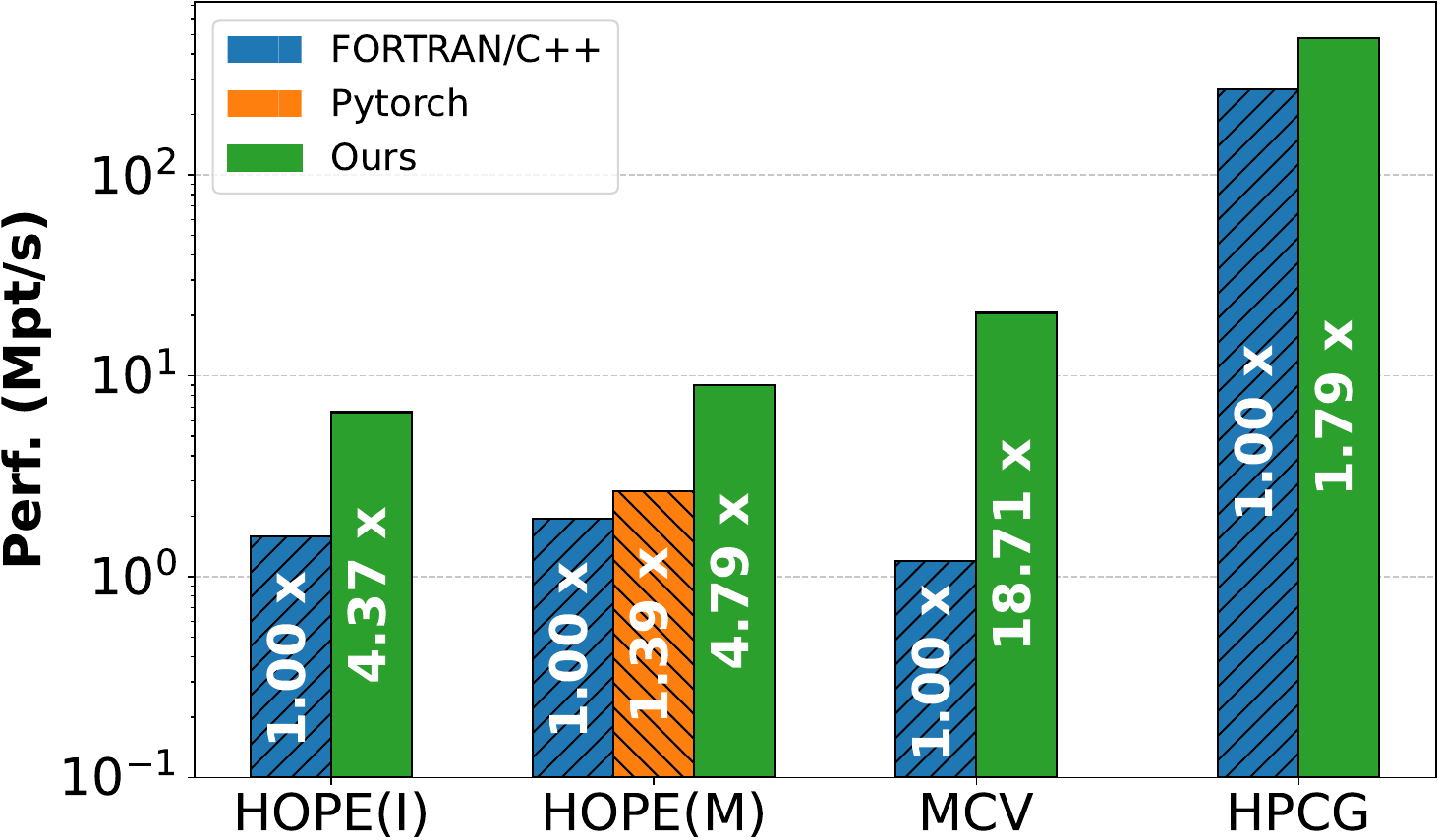}
    \caption{single node with 56 cores}
    \label{fig:eval-standalone}
    \end{subfigure}
\caption{SWE Solvers and HPCG End-to-End benchmarks}
\label{fig:eval-e2e}
\end{figure}

\Cref{fig:eval-e2e} presents the end-to-end performance of all evaluated applications using both a single core and a fully subscribed 56-core node. 
Crucially, transitioning these applications to a distributed execution model across the entire node required minimal developer effort; by modifying fewer than 20 lines of distribution configurations within the kernel codes, we successfully deployed the fully distributed versions.

As illustrated in the benchmark results, \frameworkname{} achieves an average end-to-end speedup of 1.90 $\times$ on a single core , which remarkably scales to an average speedup of 4.75 $\times$ across the fully subscribed node. Notably, while some applications like HPCG exhibit a slight single-thread overhead, they achieve massive performance gains at scale (1.79 $\times$ on 56 cores). This amplified performance at scale is primarily attributed to two architectural advantages:
\begin{enumerate}
\item The underlying designs of \frameworkname{} and the \plainmtos{} DSL exhibit superior parallel efficiency and reduced scheduling overhead in multi-threading environments.
\item Although multi-process execution introduces inherent data copying overheads, it effectively mitigates Non-Uniform Memory Access (NUMA) bottlenecks and significantly enhances overall data locality.
\end{enumerate}

\subsection{Scalability}

\begin{figure}[ht]
\centering
    \begin{subfigure}[t]{\linewidth}
    \centering
        \includegraphics[width=0.9\textwidth]{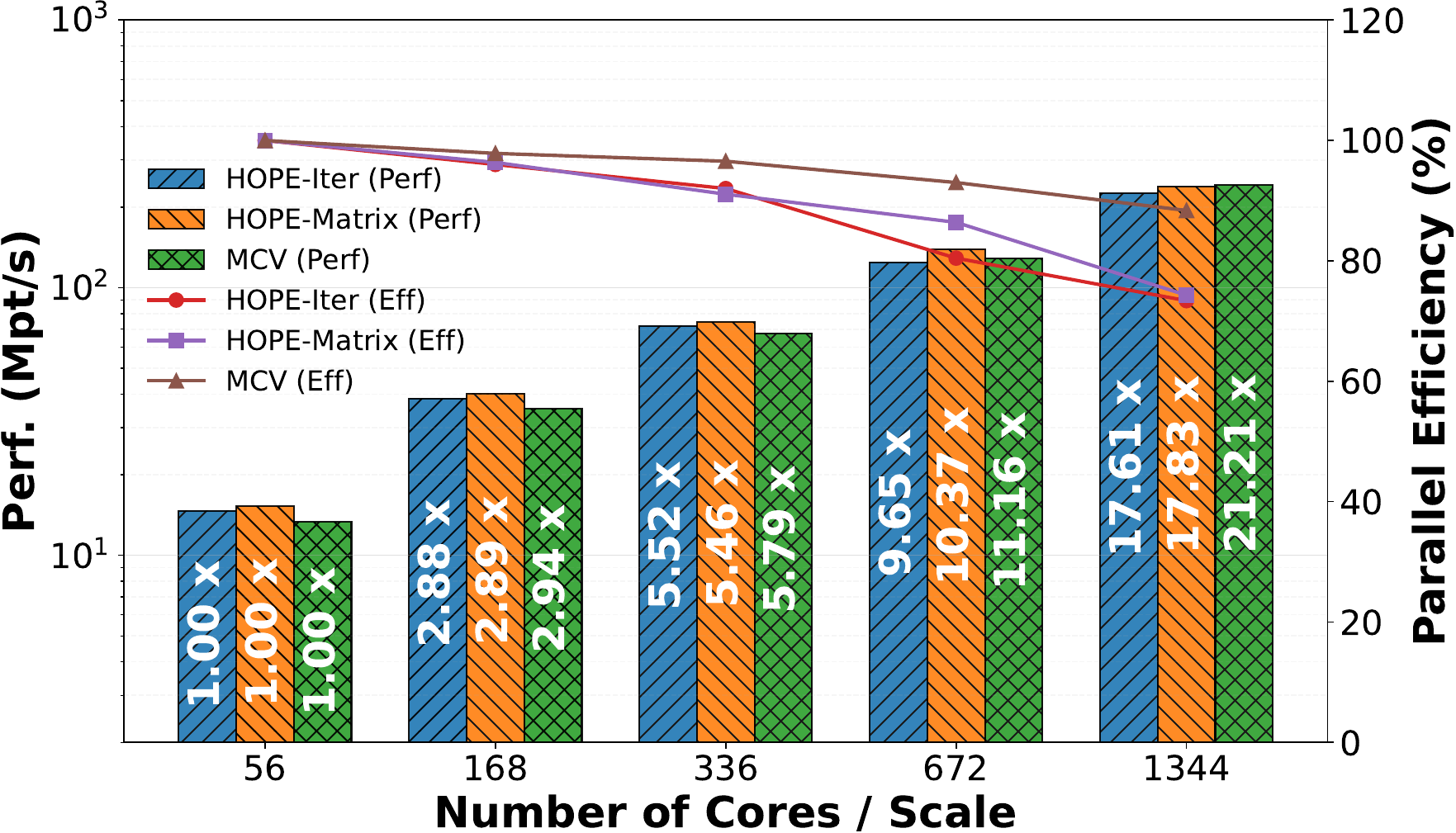}
        \caption{SWE with $6\times1440\times1440$ resolution}
        \label{fig:eval-strong-swe}
    \end{subfigure}
    \begin{subfigure}[t]{\linewidth}
    \centering
        \includegraphics[width=0.9\textwidth]{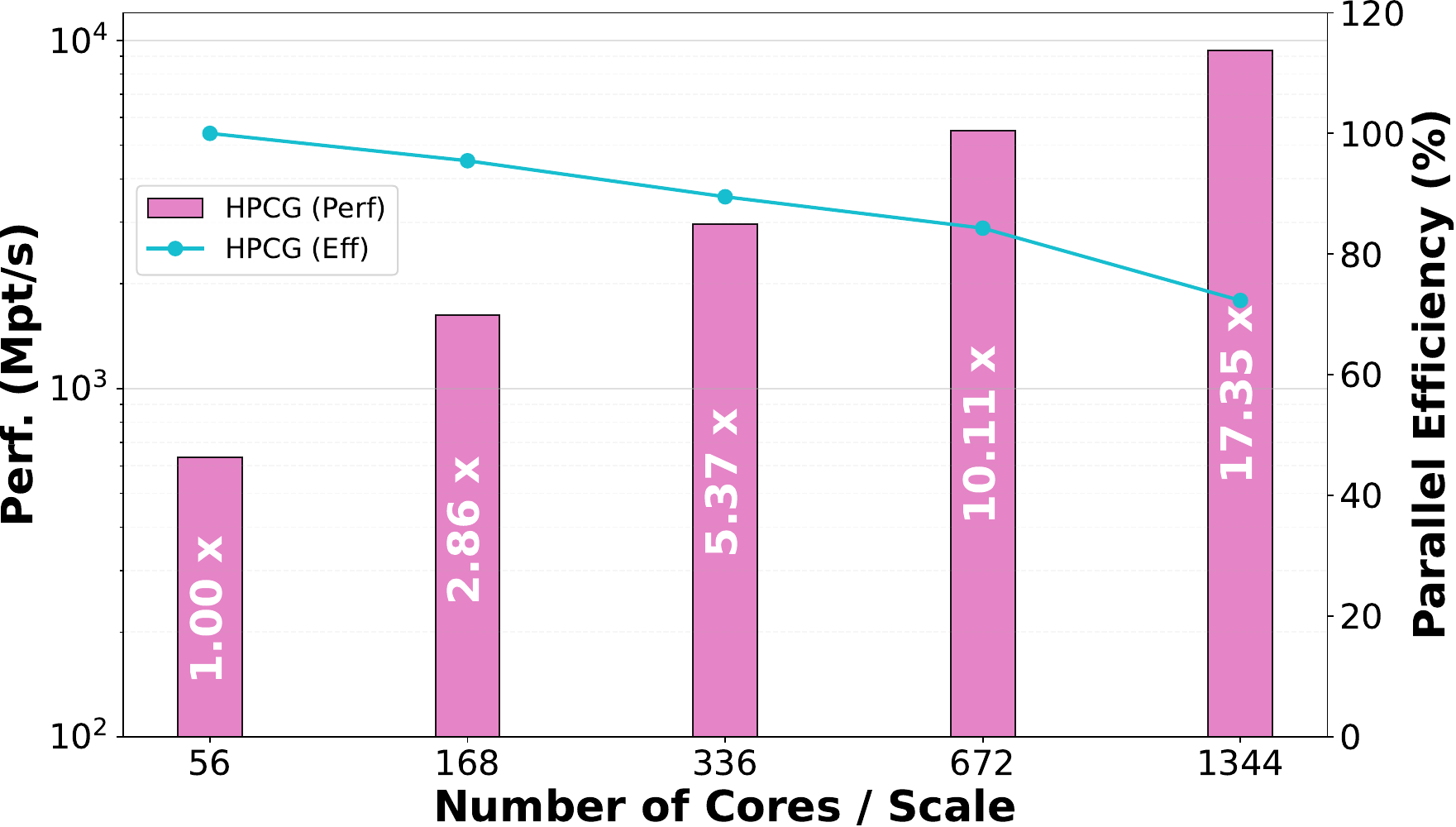}
        \caption{HPCG with $1280^3$ grid size}
        \label{fig:eval-strong-hpcg}
    \end{subfigure}
\caption{Strong Scaling Performance and Parallel Efficiency}
\label{fig:eval-strong}
\end{figure}

\Cref{fig:eval-strong} illustrates the strong scaling efficiency of our applications as we scale from a single node (56 cores) to 1,344 CPU cores (a $24\times$ scale-up). 

The applications exhibit distinct scaling behaviors based on their numerical properties. The MCV application achieves a remarkable 88\% parallel efficiency (a $21.21\times$ speedup). This optimal scaling occurs because MCV features the lowest ratio of boundary condition (BC) complexity to inner calculation workload, making it relatively compute-bound. In contrast, the higher-order HOPE methods (HOPE-Iter and HOPE-Matrix) and the communication-heavy HPCG benchmark maintain parallel efficiencies between 72\% and 74\%. For the HOPE variants, the higher numerical accuracy inherently requires thicker halo exchanges and more complex BCs, which increases inter-node communication at scale. 

These scaling results reveal a fundamental trade-off for HPC applications: users must balance the desired numerical accuracy (which dictates BC complexity) against the achievable parallel efficiency. \frameworkname{} effectively supports this design space, keeping framework-level overheads minimal even for communication-heavy workloads like HPCG.

\subsection{Evaluating Compile time}
\label{sec:eval-c-time}
\begin{table}[h]
    \centering
    \begin{tabu}{c|
    S[detect-weight, mode=text, table-format=5.2]
    S[detect-weight, mode=text, table-format=5.2]
    S[detect-weight, mode=text, table-format=5.2]
    S[detect-weight, mode=text, table-format=5.2]
    }
        {\textbf{Stages}}    & \textbf{HOPE(I)}   & \textbf{HOPE(M)}   & \textbf{MCV}   & \textbf{HPCG} \\
        \hline
         Stage 1 Execution             & 8.99            & 9.67            & 12.77            & 7.21        \\
         Stage 2 Optimize              & 648.04            & 287.40          & 273.65            & 111.36         \\
        \hline 
         Specialization & 14.28             & 11.18            & 6.09             & 462.18         \\
         GCC  & 43.55             & 39.26             & 29.66             &  33.77      \\
        \hline
        {Run}& 4618.37          & 4372.42           & 4299.07           & 906.10  \\
    \end{tabu}
    \caption{Comparing the compile and running time (in seconds) of \frameworkname{}-based programs. The running times listed are at the maximum tested scale. }
    \label{tab:compile-time}
\end{table}

To assess the compilation overhead of \frameworkname, we report the end-to-end compilation times for all workloads at their maximum scale (largest problem sizes and core allocations), as detailed in \Cref{tab:compile-time}. Thanks to our \textit{IR-reusing} technique, the stage-1 and stage-2 optimizations are executed strictly once on a 28-core NUMA of a node. For SWE solvers, this approach confines over 85\% of the compilation cost to the initial run. For HPCG, the compilation time is instead dominated by the later specialization stage due to the large number of loops inherent in multi-grid methods. Ultimately, for long-running real-world applications, this compilation overhead will finally become negligible, confirming the practicality of \frameworkname.

\section{Related Works}
\subsection{Frameworks and DSLs for Block-Structured Grids}
While numerous frameworks support block-structured grids, their handling of boundary conditions (BCs) remains a bottleneck. PETSc~\cite{PETSc}, Hypre~\cite{falgout2002hypre}, and ExaStencils~\cite{exastencils-old, lengauer2020exastencils} offer semi-structured BC interfaces, but are fundamentally tailored for linear solvers rather than Earth system models. 

Frameworks like Stella~\cite{stella}, Dawn~\cite{dawn}, GT4Py~\cite{ben2022productive, paredes2023gt4py}, and PSyclone~\cite{psyclone_bench,psyclone} target weather applications on cubed-sphere grids. However, their reliance on external BC libraries for distributed partitioning tightly couples them to specific discretizations (e.g., FV3~\cite{fv3-boundary} or GungHo~\cite{GungHo}), limiting their extensibility. 

Other tools prioritize structured inner computations over BC flexibility: Devito~\cite{devito, bisbas2025automated} supports unaligned sparse data, Open-Earth Compiler~\cite{open_earth_compiler} targets GPU execution, and Mat2Stencil~\cite{cao2023mat2stencil} simplifies explicit and implicit numerical methods. Even recent MLIR-based~\cite{MLIR} compiler efforts~\cite{compile_shared_stack} that successfully unify inner-structured calculations still lack a high-level, unified abstraction for complex boundary conditions.

\subsection{Multi-Stage Programming}
Lightweight Modular Staging (LMS)\cite{LMS} is one of such frameworks with Scala as its host language, which separates the two stages through only the type definitions: \texttt{Rep[T]} representing a \texttt{T}-typed value at the second stage. 
Mat2Stencil~\cite{cao2023mat2stencil} provides a more flexible implementation in the Python host language. 
The backtracking-and-discussing procedure is the basis of Mat2Stencil's ability to represent boundary calculations on standalone structural grids, and we extend this procedure to generate SpMV code given a list of \textit{basic sub-matrices/vectors}. 

\subsection{Polyhedral Analysis}
Extensive research exists on fine-grained dependence analysis at the statement-level granularity and its application to program transformation. Early studies are summarized in~\cite{padua1986advanced} and~\cite{kennedy2001optimizing}. 
Subsequent advances introduced a mathematical theory named polyhedral analysis for systematic dependence analysis. 
When memory access patterns are formally defined through Presburger formulas, the dependence relationship can be inferred by solving coupled equations and in-equations. 
Existing frameworks include PPL~\cite{PPL} and isl~\cite{verdoolaege2010isl}. 
Multiple compilers exploit Polyhedral Analysis and carry out optimizations on general C programs~\cite{pluto, PPCG, CHiLL} or tensor programs ~\cite{tensor_comprehension,tiramisu, tang2022freetensor}.  

We adopt FreeTensor~\cite{tang2022freetensor} as our stage 1 code generation target for its integration of isl and implementation of dependence-aware transformations. 

\section{Discussion}
While \frameworkname{} successfully modularizes boundary condition computations while maintaining high performance, several core assumptions and design trade-offs warrant further discussion:

\begin{itemize}
    \item \textbf{Necessity of Multi-Stage Programming.} 
    The multi-stage programming paradigm is fundamental to our framework. The ability to resolve specific variables at compile-time is mandatory; otherwise, the intermediate representation (IR) generated by the code generation algorithm (\Cref{sec:design-codegen}) would suffer from exponential bloat, rendering the compilation computationally intractable.

    \item \textbf{Generality of the Global Optimization.} 
    The global optimization techniques proposed in \Cref{sec:global-opt} are highly agnostic to the underlying IR. As long as the chosen IR and its infrastructure support polyhedral analysis alongside user-defined intrinsics, our optimization passes can be seamlessly applied. Once optimized, the inner-calculation kernels can be subsequently lowered into any specialized IR for more aggressive, architecture-specific optimizations.

    \item \textbf{Trade-offs in Communication-Computation Overlap.} 
    Unlike the distributed algorithms implemented in distributed Devito~\cite{bisbas2025automated}, \frameworkname{} does not explicitly overlap communications with the center-part of inner calculations. While such overlapping is technically feasible and beneficial for extreme-scale parallel efficiency in explicit methods, altering the iteration execution order breaks the semantic equivalence for implicit numerical solvers. 
    We prioritized correctness and algorithmic generality across diverse PDE solvers over marginal extreme-scale speedups.

    \item \textbf{Static Grid Assumptions and GPU Portability.} 
    Currently, the underlying communication library relies on a global initialization phase. Consequently, \frameworkname{} assumes a static grid topology and does not support dynamic meshes or Adaptive Mesh Refinement (AMR), which are occasionally used in specific computational fluid dynamics applications. Furthermore, this global initialization scheme currently presents a bottleneck for migrating \frameworkname{} to GPU architectures. Overcoming this initialization overhead for GPU portability is the primary focus of our future work.
\end{itemize}
\section{Conclusion}
We propose \frameworkname, a novel DSL that resolves the programmability and performance bottlenecks of boundary conditions (BCs) in distributed block-structured grids. By unifying diverse BCs into a sparse matrix-vector multiplication (SpMV) abstraction and integrating multi-stage programming with polyhedral compilation, our framework automatically generates highly optimized, matrix-free kernels. Evaluations on diverse workloads demonstrate profound impact: up to 7.6x speedups for BC computations, $> 70\%$ reduction in lines of code, and $72\% - 88\%$ strong-scaling efficiency across 1,344 CPU cores. 

\section*{Acknowledgement}

We use Gemini Pro AI \cite{gemini} to refine the grammar of the whole article and to generate plotting scripts for figures.

\bibliographystyle{IEEEtran}
\bibliography{citation}
\vspace{12pt}

\end{document}